\documentclass [12pt]{article}
\begin{document}

\begin{center}
  \bf{\Large{Kink radiation on small discrete periodic damped driven
      Nonlinear sine-Gordon Arrays}}
\end{center}

\vskip 0.3in

\begin{center}
 C.R. Willis
\end{center}

\vskip 0.3in

\begin{center}
  {\em Department of Physics, Boston University, 590 Commonwealth
    Avenue, Boston, Massachusetts 02215}
\end{center}

\vskip 0.75in
\begin{center}
Abstract
\end{center}

In this paper, we explain the fundamental properties of the radiation
processess of untrapped kinks moving on discrete lattices or any
spatially periodic potential. In particular we explain qualitatively
and quantitatively the interesting recent experiments of small
discrete periodic dynamical arrays of Josephson junctions. We obtain
and show the resonance condition for a kink to radiate phonons is the
kink frequency or one of its harmonics $K_{j}\dot{X}$ where $\dot{X}$
is the velocity of the kink and the $K_{j}$ are vectors of the
reciprocal lattice must equal one of the phonon frequencies,
$\omega(k)$, where $k$ are the lattice vectors determined by the
lattice spacing and the periodic boundary conditions. Then we show the
radiation resonance conditions determine the voltage at line center in
the experimental currently voltage $(I-V)$ diagrams. We also show the
phonon linewidth due to external damping and radiation damping gives
rise to steps and hysteresis in the $I-V$ diagrams. In addition in the
cases of more than one kink, say $M$ kinks, there is a simple allowed
value of $k$ in contrast to the $M=1$ case where there are at least as
many resonances as there are phonon modes. We show the unique $k_{M}$
for a given $M$ is the allowed $k$ such that the distance between the
kinks on the ring is one wavelength i.e. $\lambda
_{M}=(2\pi)k_{M}^{-1}$.

\newpage 
\section*{\large{I. INTRODUCTION}}

The purpose of this paper is to explain the fundamental properties of the 
radiation process of untrapped kinks moving on discrete lattices or any 
spacially, periodic potential. One of our principle aims is to explain the 
very interesting experiments and simulations of small discrete periodic 
dynamical arrays of Josephson junctions in a recent paper by Watanabe et al. 
[1]. In section II we first explain the radiation properties of kinks in the 
large (quasi-continuous phonon spectra) discrete sine-Gordon (SG) system 
because most of the fundamental kink radiation interactions are quite 
similar in the large and small systems. The solution of the kink radiation 
problem for the large discrete SG system by the rigorous collective variable 
(CV) theory including radiation from trapped kinks of [2] was found to be in 
excellent agreement with the simulations of the discrete SG even in the 
strongly nonperturbative regime. Unfortunately the full nonperturbative 
equations do not lead to a simple description of the radiative interactions. 
However when a perturbative treatment of the kink radiation process is valid 
we find we can understand the results in a simple intuitive way. The 
circular arrays of Josphson junction experiments are in a parameter range 
where the kink radiation interaction is treatable by perturbation theory 
between the center of the mass of the discrete SG kink $X(t)$, which satisfies 
the non linear Peirels-Nabarro (PN) equation of motion in the presence of 
the kink. The radiation phenomena that occur in the Josephson junction 
experiments and in the large discrete SG systems we analyze in this paper 
come from kinks that are not trapped in the PN but are almost free i.e., the 
kink kinetic energy is much larger than the kink potential energy.

An important early large discrete system radiation result for which 
perturbation theory is valid was that of [3] who simulated and analyzed a 
large discrete SG system and found for a variety of initial conditions that 
there were certain special velocities of the kink at which the rate of loss 
of energy due to radiation of phonons changed abruptly. They called these 
special velocities ``knees.'' They obtained numerical values for the 
velocity knees by a graphical solution of the dispersion law of the discrete 
SG which agreed with the simulations. In [2] we pointed out the velocity of 
the kink at the knees satisfied the condition $(2\pi j\dot{X}/a) \equiv 
K_j \dot{X}=\omega _0$, where $\omega _0$ is the lower band edge of 
the phonon band, $\dot{X}$ is the velocity of the center of mass of the kink,$a$ is 
the lattice spacing and $j$ is an integer. In the present paper we show the 
radiation resonance conditions, $K_j\dot{X}=\omega _0$, for the large 
system is a special case of the general radiation resonance conditions which 
play the crucial role determining the value of the voltages in the 
current-voltage characteristics in the small discrete periodic systems. The 
frequency $K_j\dot{X}$ is the $j$-th harmonic of $2\pi\dot{X}/a$, which we call the 
kink frequency, for reasons that will become clear in Sec. II. In the small 
discrete system there are a set of radiation resonance condition 
$K_{j}\dot{X}/a=\omega(k)$ where $\omega(k)=[\omega _0^2+ 
4sin^2 ka/2]^{1/2}$, $k$ is the wave vector of the kth phonon mode, and 
the $K_j$ are the reciprocal lattice vectors of our discrete lattice 
spacing of $a$. We show the large and small discrete systems radiation 
resonance conditions arise from the calculation of the radiation from a kink 
moving at approximately constant velocity, $\dot{X}$. In the large system the 
radiation resonance conditions give directly the value of the velocity knees 
$\dot{X}_j=\omega _0 K_j^{-1} a$. One of the conditions for the 
validity of perturbation theory of kink radiation is different in the large 
and small system cases. In the large system the phonon states form 
essentially a continuum of phonon states which leads to the requirement, 
$\omega _o\tau _{r}\gg 1$, where $\tau _{r}$ is the radiation 
relaxation-time. This condition is satisfied in [3] and all the large 
systems considered in this paper. While in the small systems where because 
of the large spacing between phonon frequencies it is necessary to have a 
weak external dissipation term, $\Gamma\dot{\phi} $, in the discrete SG 
equation to provide the phonon line with a linewidth which leads to the 
requirements, $\Gamma \tau _{r}\gg 1$ which is satisfied in the 
experiments. In addition to $\Gamma$ in the small system there is a 
contribution to the phonon linewidth due to spontaneous emission, $[\tau 
(k)]^{-1}$, where $\tau $ is the phonon lifetime which is a non linear 
function of $k$ and $\dot{X}$. We gain further insight into the kink radiation problem 
by showing how the continuum SG with a spacially periodic perturbation, $\cos 
k'x\;\; \phi(x,t)$, which has been solved in perturbation theory and 
has a single knee at $k'\dot{X}=\omega _0$. This example makes 
clear the importance of spatial periodicity of the potential in the kink 
radiation problem by defining the kink frequency, $k'\dot{X}$ in the 
perturbed continuum. SG and $K_j \dot{X}$ in the discrete periodic lattice case.

In Sec. III, we show the number of degrees of freedom and the periodic 
boundary conditions determine the allowed wave vectors which together with 
the radiation resonance conditions determine the velocities of the kinks at 
the phonon line centers. The voltage is proportional to $\dot{X}$ which in turn is a 
non linear function of the driver $I$. Consequently we find $I$ is a non linear 
function of $V$ which leads to the steps in the $I-V$ diagrams. The non-linear 
dependence of $I$ on $V$ in small driven damped periodic rings leads to 
hysteresis between adjacent steps.

We find the perturbative treatment of the CV equations of motion explain 
quantitatively the main experimental and numerical results of Watanabe et 
al. [1] on rings of Josephson junctions who consider rings of $N=4$ and $N=8$ 
junctions with $M=1,2,3$ and 4 kinks for various values of the discreteness 
parameter. In all cases the perturbation theory is valid and agrees with the 
experiments and simulations to within three percent. In addition in the 
cases of more than one kink, say $M$ kinks, there is a simple allowed value of 
$k$ in contrast to the $M=1$ case where there are at least as many resonances as 
there are phonon modes. We show the unique $k_M$ for a given $M$ is the 
allowed $k$ such that the distance between the kinks on the ring is one 
wavelength i.e. $\lambda _M=(2\pi)k_M^{-1}$.

\section*{\large{II. RADIATION-RESONANCE CONDITIONS IN LARGE AND SMALL SYSTEMS}}

In this section we analyze the radiation mechanism of almost freely moving 
kinks in large and small discrete lattices where the perturbative effect of 
discreteness causes the kinks to radiate phonons. The continuum SG in 
dimensionless variables is :

$$
 \ddot{\phi} -\phi''+(\pi /l_o)^2 \sin\varphi= 0  \eqno(2.1)
$$

where the dot represents time derivative, the prime represents spatial 
derivative and $(\pi /l_o)^2$ is the dimensionless constant. The 
size of the kink in these variables $\sim l_o$. The dimensionless 
parameter $\pi /l_o$ is the slope of the kink and $(\omega _o=\pi 
/l_o)$ is the frequency of the lower phonon band edge. A complete 
discussion of the units is given in [2]. The discrete SG in the same units 
with the lattice spacing $a=1$ is :

$$
 \ddot{\phi}_n-\Delta _2 \phi_n +(\pi/l_o)^2 \sin\phi _n=0  \eqno(2.2)
$$
where
$$  
 \Delta _2 A_n = A_{n + 1}+ A_{n - 1}-2 A_n            
$$

The values $(\pi /l_o) >1$ are considered discrete and the continuum 
limit corresponds to $\pi /l_{o} \ll 1$. We assume a solution in the form 
$\phi _n=f_n+q_n$ where $f_n=4\tan^{-1}\exp[(\pi /l_o 
)(n-X(t))]$ and $\Sigma _2 q_j f'_j=0$ and $\sum 
p_j f'_j=0$ are the constraints where $f'_j$ is the 
derivative of $f_j$ with respect to $X(t)$ and is called the shape mode. The 
derivation of the complete CV equations of motion for $q_n(t)$ and $X(t)$ 
are given in [2.] In all cases we consider the kink radiation interaction 
can be treated perturbatively. Consequently we need only the lowest order 
theory namely the nonlinear equation for the center of the mass $X(t)$ in the 
Peirels-Nabarro PN potential and the linear equations for the phonons 
$q_n$ on the presence of the kink. The lowest order equation of motion for 
$X(t)$ is 

$$
M(X)\ddot{X}= -\sum_n f'_n [\Delta _2 f_n-(\pi 
/l_o)^2 \sin f_n]+\frac{1}{2} \dot{X}^2 \frac{d}{dX}\sum_n(f'_n)^2
$$
and terms linear in q see Eq. (2.11) of [2], where $M(x)\equiv \sum ( 
f'_n )^2=M_{o}+M(X)=8(\pi/l_{o})+\Delta M(X)$.

For the levels of discreteness in the experiments and simulations we analyze 
in this paper $\Delta M(X)\ll M_{o}$ thus we can replace $M(X)$ by
$M_{o}$ and we 
can neglect the $\dot{X}^2$ term. The linear terms in $q$ in the $X$ 
equation represent induced absorption and emission terms which are 
negligible compared with the spontaneous emission term in the experiments 
and simulations analyzed in this paper. Finally we obtain 

$$
 \ddot{X} = -( M_o)^{-1}\sum _n f'_n[\Delta _{2}f_{n} -(\pi/ l_{o})^{2}\sin f_{n}]  
   \equiv \sum _j C_j\sin 2 \pi j X(t)/a                           \eqno(2.3)
$$

The dominant term in $C_{j}$ goes as $C_{j}\exp(-j \pi l_{o})$. Thus 
$C_{j} $ decreases with increasing $j$ and $l_{o}$. Thus as the size of the 
kink, $2l_{o}$, increases compared with $a$, $C_{j}\to 0$ and in 
this limit of the continuum the force due to discreteness vanishes. For 
appreciable discreteness effects the size of the kink should be less then 
six or seven lattice spacings. The linearized equations of motion for the 
$q_{n}$ are:

\begin{eqnarray*}
\ddot{q}_n - \Delta_2 q_n + [(\pi /l_{o})^{2}\cos 2 \pi f_{n}]q_{n}&=&\ddot{X} f'_{n} +\Delta_{2}f_{n} 
-(\pi /l_{o})^{2}\sin 2\pi f_{n}  \\ 
&\equiv& \sum _{j} d_{j}(n) \sin 2 \pi jX/a   \hskip 1.0in (2.4)
\end{eqnarray*}

The closed coupled set of equations for $\ddot{X}$ and $\ddot{q}_{n}$ Eqs. (2.3) and 
(2.4) constitute the lowest order CV equations which explain the spontaneous 
emission of a SG kink in a discrete lattice. The normal modes and 
eigenvalues for the $q_{n}$ when the right hand side of Eqs. (2.3) and (2.4) 
are set equal to zero are
$$
 \omega ^{2} (k) = (\pi /l_{o})^{2} + 4\sin^{2}(k/2)
$$
 and 
$$
\psi_{k}(n) = [2\pi \omega (k)]^{-1} \exp [k\pi n/l_{o}]\{ik  
- \tanh[\pi n/l_{o}]\}                                      \eqno(2.5)
$$
The phonon modes in the presence of the kink, Eq. (2.5) are orthogonal to 
the shape mode of the kink $f'_{n}$. Note the spectra $\omega(k)$ 
is the same whether the kink is present or not. Strictly speaking Eq. (2.5) 
and $\omega(k)$ are rigorously true in the continuum limit and 
approximately true in the weakly and moderately discrete region. All of the 
experiments and simulations we consider in this paper are in the moderately 
and weakly discrete limit. In the extreme discrete limit $\pi /l_{o}>1$ 
where the kink is only two or three lattice spacings the eigenfunction and 
eigenvalues are not known analytically but only numerically because in the 
extreme discrete limit the ground state is only known numerically. We thus 
have a model of a point ``particle'' $X(t)$ with Hamiltonian
$$
  H_{\kappa}  = \frac{P_X^2}{2 M_X} + V_{PN}(X)  \eqno(2.6)
$$
where
$$ 
V_{PN}(X) \equiv \sum 
C_{j}(2\pi j)^{-1}[1-\cos 2 \pi j X/a] 
$$

spontaneously emitting phonons through the terms on the right hand side of 
Eq. (2.4). The PN frequency $\omega _{PN}^{2}=\sum 
C_{j}^{2} 2 \pi j$. From perturbation theory the energy of the kink plus 
the energy of the phonons is conserved. There are two kinds of spontaneous 
emission of phonons. The first comes from a kink trapped in a PN well, i.e., 
when $\frac{P_ X^{2}}{M_{X}}<V_{PN}(X)$ and the 
kink radiates harmonics of $\omega _{PN}$ where $m\omega _{PN}>\omega 
_{o}$ and $m$ is an integer. This is anharmonic radiation which is treated 
in detail in [2] and plays no role in the phenomena analyzed in this paper. 
The phonon radiation mechanism which explains the phenomena analyzed in this 
paper is determined by the frequency which the center of mass of the kink, 
$X(t)$, passes the lattice points separated by a distance a apart when 
the kink velocity is approximately constant which is $\Omega =2\pi  
\dot{X}/a\equiv K\dot{X}$, which we call the kink frequency, and its 
harmonics $\Omega _{j}=2 \pi j\dot{X}/a \equiv  K_{j}\dot{ 
X}$. Which can also be seen by setting $ X(t) \approx \dot{X}t$ in 
$\sin 2 \pi j X(t)/a$ which becomes $\sin 2 \pi (K_{j}\dot{X})t 
\equiv \sin 2 \pi \Omega _{j}t$. Note this frequency is determined only 
by the periodicity of the lattice and does not depend on the strength of 
interaction as $\omega _{PN} $ does. The $K_{j} $ are the vectors of 
the reciprocal lattice. For radiation of frequency $\Omega $ the kink has to 
be essentially free in the sense that the kinetic energy is appreciably 
greater than the potential energy i.e., $\frac{P_ X^{2}}{2M_X}\gg 
V_{PN}$. In order to see the role played by 
$\Omega $ in kink radiation processes consider the solution for 
$$
 \phi_{n}(t)=\sum \psi _{k}(n) \exp [-i \omega(k)t]\sum 
_{j}C_{jk}\int_{0}^{t} dt' \exp i[\omega (k)t'-2 \pi j\dot{X}(t')/a]  \eqno(2.7)
$$
where 
$$
 C_{jk} \equiv \sum _{n}d_{j}(n) \psi^* _{k}(n) .
$$
Now if we can treat the kink as a free particle for times t small compared with 
the radiation relaxation time but long with a suitably short interaction 
time we can replace $X(t)$ by $\dot{X}t$ in Eq. (2.7) and obtain the usual delta 
function of perturbation theory exactly as in Fermi's golden rule. In the 
large system the phonon levels become continuous and the interaction time is 
$\omega _{0}^{- 1}$. When we let $t \to \infty $ in Eq. (2.7) we 
obtain 
\begin{eqnarray*}
\phi(n,t)&=& \int dk \psi _{k}(n)\exp[-i \omega(k)t]\sum 
_{j} C_{jk}\;\delta [\omega(k)-2 \pi j\dot{X}/a] \\
 &=&\sum _{j}\psi 
_{k(j)}(n)\exp[-i 2 \pi j\dot{X}/a]\;t\; C_{jk(j)}\;\omega [k(j)]/k(j)
\hskip 1in (2.8)
\end{eqnarray*}

where $k(j)$ is the solution for $k$ of the radiation resonance condition 
$\Omega _{j}=\omega [k(j)]$, and where for convenience we have taken 
units where $\pi /l_{0}=1$. Thus for a given $j$, say $j=1$, our solution, 
$\phi(n,t)$ is a single normal mode $\psi _{k(1)}(n)$ with a single 
frequency $2\pi\dot{X}/a$ which is the phonon frequency radiated by the kink 
which is resonant with the kink frequency $K_{j}\dot{X}$. The general solution is 
a linear combination of those phonons in the phonon band which are resonant 
with the kink frequency and its harmonics $K_{j}\dot{X}$. The damped driven small 
system does not have a delta function radiation resonance condition but a 
Lorentzian centered on the resonance condition with a finite linewidth. We 
discuss the effect of the finite linewidth on Secs. III and IV. The apparent 
divergence of Eq.(2.8) at $k=0$ is spurious and is not a breakdown of 
perturbation theory. It is a breakdown of the assumption that $\dot{X}(t)$ is a 
constant which is true everywhere except at the band edge where the time 
dependence of $\dot{X}(t)$ has to be considered within perturbation theory. For a 
full discussion of the treatment of $\dot{X}(t)$ in the large system see [4]. The 
spurious divergence does not arise in the small discrete system because of 
the external damping $\Gamma\dot{\phi} _{n}$.

For each phonon mode, $\omega(k)$, there will be a set of velocities $\dot{X}$ such 
that $\Omega _{j}=\omega(k)$. The ratio of successive kink velocities 
resonant with a given $\omega(k)$ is $(X_{j + 1}/X_{j})=(j/j+1)$. The 
number of harmonics $\Omega _{j}$ and the number of kink velocities that 
correspond to a given phonon mode, $\omega(k)$ is determined by the 
discreteness parameter $(\pi /l_{0})$. As the system becomes more discrete 
i.e., as $l_{0}$ decreases the more $C_{j}$ will contribute to Eq. (2.3) for 
$V_{PN}$ resulting in more harmonics $\Omega _{j}$ contributing to the 
radiation by the kink. However as the system becomes more discrete the PN 
potential well will become deeper and the kink becomes trapped sooner, the 
range of velocities for which $\frac{P_{x}^{2}}{2 M _{x}}\gg V_{PN}$ decrease 
leading to an upper limit on the j values such that the ``free condition'' 
is valid. When $\Omega _{j}$ lies in the phonon band the kink loses energy 
by radiation of phonons. The unperturbed energy of the kink plus phonons is 
conserved. When the velocity of the kink decays to $X_{j}=(2\pi j)^{ - 
1}\omega _{0}$ the kink abruptly stops radiating at the frequency 
$\Omega _{j}$ because there are no phonon states $\omega(k)<\omega 
_{0}$. This sharp decrease in the rate of loss of energy by the kink is 
called a knee in Ref. [3]. The remaining radiation comes form higher 
harmonics $\Omega _{j'} =2 \pi j'\dot{X}/a$ where $j'>j$ 
which are still in the band and still radiating. When the velocity of the 
kink decays further to $\dot{X}_{j + 1}=[2\pi (j+1)]^{-1}\omega _{0 
}$ the kink stops radiating the j+1 harmonic causing the next knee in the 
radiation rate. The pattern is repeated until the kink loses enough energy 
to be trapped in a PN well where it will radiate the harmonics of $\omega 
_{PN} $ at a much lower rate.

Peyrard and Kruskal [3] simulated and analyzed the discrete SG for a large 
system and found for a variety of parameter values $(\pi /l_{0}=.9,1,$ and 
1.05) and initial values of the velocity that there were one, two or three 
knees. They obtained numerical values for the velocity knees by a graphical 
solution of the discrete SG dispersion law which agreed with their 
simulations to better than one percent. Their numerical values for the three 
knees are given by our radiation resonance condition for $j=1, 2$ and 3 to 
within one percent. The velocity of the jth knee is given by $\dot{X}_{j}=(2\pi 
j)^{-1}\omega _{0}$ as was pointed out in [2].

There is one continuum SG kink radiation problem that gives further insight 
into the general kink radiation problem and that is the model of the 
continuum SG with a spatially periodic perturbation potential which was 
analyzed in Ref. [4]. The perturbed SG equation is 
$$
\ddot{\phi} - \phi'' + (\pi /l_{0})^{2}(1+\epsilon\sin \;\;k'x) \sin \phi=0
$$
where $\epsilon $ is a small dimensionless parameter and $k'$ is 
the wave number of the perturbation. The competing length scales are 
$l_{0}^{-1}$ and a i.e., the size of the kink and the lattice spacing. 
When we go to a CV description [4] by making the ansatz $\phi =\sigma 
[(\pi /l_{0})(x-X(t))]+\chi [(\pi /l_{0})(x- X(t)),t]$ where $\sigma $ 
is the kink solution of Eq. (2.8) when 
$\epsilon=0$ and to lowest order in $\epsilon $ we obtain the 
following coupled equations of motion 
$$
 \ddot{\chi} - \chi '' + (\pi /l_{0})^{2} \cos\sigma \chi = -\epsilon
 (\pi /l_{0})^{2} \sin k' X
$$
 and 
$$
\ddot{X}+\omega^{2} \sin k'X=0
$$
where 
$$
\omega ^{2}=(\epsilon /l_{0})(k'l_{0}/2)^{2}[\sin(k' l_{0}/2)]^{ - 1}.
$$

The equation for $X$ is the pendulum equation. In[4] the radiation and
$X(t)$ are calculated analytically. The relevance of this example for
the present discrete case is that the $\ddot{X}$ equation is of the
same form as Eq. (2.3) with $C_{1}=\omega^{2}$ and $C_{j}=0$ for $j
\ge 2$.  Consequently only one velocity knee occurs i.e., when
$=\dot{X} = (\omega _{0}/k')$ which is the same as our radiation
resonance result for $K_{1}=k' $. The kink radiates phonons at
frequency $\omega(k)$ and wave number $k$ such that
$k'\dot{X}=\omega(k)$. The example of the continuum SG kink in the
$(\epsilon \sin k'x)\phi$ perturbation makes clear that the kink
frequency which is resonant with the phonon frequency $\omega(k)$ when
$X(t)\sim \dot{X}t$ is determined by the spatial periodicity of the
perturbing potential which is $K_{j}^{-1}$ in the discrete case and
$(k')^{-1}$ in the perturbed continuum case.  The resonance condition
arises in the same way in both cases i.e., $\delta
[\omega(k)-k'\dot{X}]$ and $\delta[\omega(k)-K_{j}\dot{X}]$ after
setting $X(t)\approx \dot{X}t$. We next consider the kink in small
finite systems which satisfy periodic boundary conditions as in the
experiments with Josephson junction rings, [1], where cases of 4 and 8
degrees of freedom with 1,2,3 or 4 kinks are treated. In the small
discrete periodic systems the kink interacts with the radiation it has
spontaneously emitted. However, the induced absorption and emission
processes are much smaller than the dominant spontaneous emission
process and don't play a significant role.  They have the same
radiation resonance condition as the spontaneous emission process. One
difference appears in the small periodic system is the phenomena of
hysteresis which we discuss in Sec. IV.

In the Josephson junction rings experiments the kink radiation
interactions are perturbative which implies the phonon eigenmodes and
eigenvalues are those of the linear phonon equations in the presence
of the kink. Note the number of modes and the spectra $\omega(k)$ are
unaffected by the presence of the kink. Since many discussions of the
experiments often have an incorrect number of modes and frequencies of
the modes it is worth while to provide a brief elementary review [5].
The phonon modes in a lattice of spacing a with $N$ particles
satisfying periodic boundary conditions have displacements $u(sa)$
where $s$ is an integer which satisfy $u[sa] = u[sa+L] = u[(s+N)a]$
where $L=Na$. The solution for $u$ is $u[sa,t] = u[0] e^{i[ska -
  \omega (k)t]}$ where $k=0,\pm 2\pi/L,\pm 4\pi/L \cdots N\pi/L$ and
$\omega(k) = [1+4 \sin^2 (ka/2)]^{1/2}$ and where for convenience we
use units where $(\pi/l_{o})= 1$.

The number of phonon modes is equal to the number of particles N which
is also the number of $k$ values i.e., $k=0, \pm \cdots \pm (N-2)\pi
/L, N\pi/L$ where the modes $k=0$ and $N\pi/L$ are non degenerate and
the remaining modes are two fold degenerate with one mode traveling in
the clockwise direction and one in the counterclockwise direction with
the same frequency $\omega(k) = \omega(-k)$. The $k=0$ mode
corresponds to all $N$ particles oscillating in phase at the frequency
$\omega _{0}$. The $k=\pi $ corresponding to $\lambda=2$, where each
particle oscillates 180 degrees out of phase with its nearest neighbor
at a frequency of $(\omega _{0}^{2}+4)^{1/2}$. The number of different
phonon frequencies $\omega(k)$ is $[(N-2)/2 +2] = (N/2) +1$. For a
typical case of $N=8$ there are five different frequencies
$\omega(0)$, $\omega(\pi/4)$, $\omega(\pi /2)$, $\omega(3\pi/4)$, and
$\omega(\pi)$ where we set $a=1$. The crucial point is that in the
radiation resonance condition,$ K_{j}X = \omega(k)$, that the $K_{j}$
are vectors of the reciprocal lattice and are completely different
than the vectors $k$ that determine the phonon modes of the lattice.
As an illustration consider the case $N=8$, $j=1$ and $M=1$. For each
of the five different values $\omega(k)$ there will be one velocity of
the kink which is resonant namely, $\dot{X}(k) = a(2\pi
)^{-1}\omega(k)$ for a total of five different velocities. The next
question that arises is given the number of phonon frequencies is
fixed by $N$ i.e., $(N/2)+1$, how many different kink velocities can
be resonant with the five modes. The answer is determined by the
number of $j$ values that contribute to $V_{PN}$ without the kink
being trapped. For each $j$ value there are $[(N/2)+1]$ velocities so
the answer is $j[(N/2)+1]$ different values of $X$ and thus the same
number of voltages would be possible. The number of $j$ values is
determined by the discreteness parameter $\pi/l_{o}$.

There is a theorem [6] which relates the total number of degrees of
freedom to the number of kinks and phonons which states the total
number of degrees of freedom $N$ must satisfy $N=N_{K}+\int\rho(\omega
)d\omega$ where $N_{K}$ is the number of kinks and $\int\rho (\omega
)d\omega$ is the effective number of phonon degrees of freedom with
$\rho(\omega)$ the phonon density of states. This theorem does not
reduce the number of eigenfrequencies, $\omega _{\alpha}$, but their
effective weight i.e., in our case $\rho (\omega )=\sum _{\alpha}
C_{\alpha}\delta(\omega - \omega _{\alpha} )$ thus we have $N =
N_{K}+\sum C_{\alpha} $. The states $\omega _{\alpha} $ are the same
as in the absence of the kink but the effective number of phonon
states is $\sum C_{\alpha} = N - N_{K}$ reduced from the $N$ it would
be in the absence of kinks.

In this section we have seen the radiation resonance condition at line
center, $K_{j}\dot{X} =\omega(k)$ is the same in the large discrete
periodic system. The number of $j$ values i.e. harmonics of $2\pi\dot{X}/a$
is determined by the boundary conditions and the dispersion law for
$\omega(k)$ is the same in both cases. The small periodic system
requires external damping to justify radiation perturbation theory
while in the infinite system the quasi continous density of states is
sufficient. We defer to Sec. IV a discussion of hysteresis which
occurs in the small periodic system and for which there is no analogue
in the large system.

\section*{\large{III. ANALYSIS OF JOSEPHSON JUNCTION RING EXPERIMENTS}}

In a recent paper Watanabe et al. [1] presented an analyzed
experimental measurements and simulation results for the
current-voltage (IV) characteristics of rings of 4 and 8 underdamped
Josephson junctions using a model of the damped driven SG equation
with periodic boundary conditions.  The SG equation in their notation
is $\ddot{\Phi}_{n}+\Gamma \Phi _{n} + \sin\Phi _{n} -\Lambda
^{2}\Delta _{2}\Phi _{n}= I/I_c$ where $\Gamma $ is the damping and
$I/I_c$ is the damped driven form of Eq.  (2.2) the dimensionless
parameter $\Lambda ^{2}$ is associated with the $\Delta _{2}$ term
while in Eq.  (2.2) the parameter is associated with the nonlinear
$\sin\phi _{n}$ term.  They found two types of traveling waves low
velocity kinks and high velocity whirling modes.  In this paper we are
concerned only with the low velocity kink modes.  Their analysis of
the behavior of the low velocity kinks in their words ``consists of
rough physical arguments that are far from rigorous.'' In this section
we show the CV radiation perturbation theory in Sec. II correctly
explains the number of peaks and their position in the I-V graphs very
accurately. In particular the experimental peaks are determined by the
phonon normal modes, the kink frequencies, $2\pi j\dot{X}/a$, the
radiation resonance, condition, $2\pi j\dot{X}/a =\omega(k)$, as
presented in Sec. II.

In order to explain the I-V diagrams we first review the voltage phase
relationship as given in Eq. (2.16) of [1] as the Josephson
relationship $V_{j}= V_{0}\dot{\Phi} _{j}/\Lambda$ where for a kink
$\dot{\Phi} _{j}=\Phi _{j}'\dot{X}$ which shows for a kink traveling
with a constant velocity that $V_{j}$ for each $j$ is $\sim X$. For a
phonon normal mode all the $\dot{\Phi} _{j} \sim\omega(k)$ and thus we
have $V_{j}\sim\omega(k)$. For a kink moving at constant velocity
$\dot{X} = a/T$, where $T$ is the time to travel the lattice spacing
$a$, we have the kink frequency $\omega= 2\pi\dot{X}/a$ in the
dimensionless units of Eq.(2.2). Thus the space and time average of
$V_{j}/V_{0}\equiv V/V_{0}$ for a constant velocity kink is $V/V_{0} =
2\pi\dot{X}/a$ or $V/V_{0}=\omega/\Lambda $ in the dimensionless units
of [2.2]. In the case of a phonon frequency we have $V/V_0 =
\omega(k)/\Lambda \equiv [\Lambda ^{- 2} + 4 \sin^{2}(ka/2)]^{1/2}=
[\omega _{o}^{2}+ 4 \sin^{2} (ka/2)]^{1/2}$ where in the last equality
we used the relationship between the two dimensionless set of
variables of [2.2] and [3.1] where $(\pi/l_{o})^{2}=\Lambda ^{ - 2}$.
However because of the radiation resonance condition the frequency of
the phonon and thus $V/V_{0}= 2\pi \dot{X}/a = [\omega
_{0}^{2}+4\sin^{2} (ka/2)]^{1/2}$ which is true independent of space
time averaging.

Thus the line center of each phonon line, $\omega(k)$, determines a
unique value of the voltage $V$. Since $\dot{X}$ is a nonlinear
function of $I$ we have an $I(V)$ for each phonon line. We show how
the finite phonon linewidth in a small discrete periodic systems leads
to the full $I(V)$ lines and hysteresis later in this and the next
section.  We also point out that the preceding conclusion for
$V/V_{0}= 2\pi \dot{X}/a =\omega(k)$ is valid also for the harmonics
of the kink frequency i.e.  $V/V_{0}= 2\pi j\dot{X}/a =\omega(k)$ for
$j$ an integer as long as for those $j$ values the kink is untrapped
as discussed in Sec. II.

In order to avoid all effects of space time averaging I compare the
ratio of the experimental voltages to the ratio of the corresponding
phonon frequencies and thus to the ratio of the kink velocities. I now
show the CV kink-radiation theory of [2,3] determines the number and
position of the voltage lines in the I-V diagrams for all the
experiments and simulations including new results for the multi kink
cases. We consider the individual cases one at a time. The first case
we consider is a single kink, $M=1$, $N=8$ junctions, discreteness
parameter $\Lambda=1.48$ and $\Gamma =0.13$ in fig.10 of [1]. In fig.
10 there are five peaks in the I-V diagram determined by the five
minima of the $\frac{d(V/V_{c})}{d(I/I_{c})}$ curve which corresponds to the
correct five phonon frequencies for $N=8$ junctions with periodic
boundary conditions. (On Fig. 10 a sixth point is indicated to be a
minimum but does not appear to be from the figure.) We next show that
the measured voltage ratios agree with the corresponding, phonon
frequency ratios to about $2.5{\%}$. The ratio we calculate are
$$
\left|\omega(k+1)/\omega(k)-V(k+1)/V(k)\over \omega(k+1)/\omega(k)\right|
\eqno(3.2)
$$
where the $\omega(k)$ are given by Eq.(2.5) for $k=0, \pm
\pi/4,\pm\pi/2,\pm3\pi/4,\pi$ and the voltage ratios are obtained 
from the experimental values given in Fig. 10. The ratios of the
succesive calculated phonon frequencies are 
$$
\frac{\omega(\pi/4)}{\omega(0)}= 1.513, \;\;\;\;
\frac{\omega(\pi/2)}{\omega(\pi/4)}= 1.54, \;\;\;\;
\frac{\omega(3\pi/4)}{\omega(\pi/2)} = 1.26,\;\;\;\;
\frac{\omega(\pi)}{\omega(3\pi/4)} = 1.07
$$
while the corresponding experimental values of the four voltages are
$$
\frac{V(1)}{V(0)}=1.47, \;\;\;\;
\frac{V(2)}{V(1)}=1.54, \;\;\;\;
\frac{V(3)}{V(2)}=1.29, \;\;\;\;
\frac{V(4)}{V(3)}=1.107 \;\;\;\;
$$
where $V(0)$ is the voltage corresponding to $\omega(k)$ for $k=0$
and the minima of $d(V/V_{c})\over d(I/I_{c})$, $V(1)$ is the second
minima of $d(V/V_{0})\over d(I/I_{0})$ corresponding to
$\omega(\pi/4)$ etc. When we substitute the phonon frequencies
$\omega(k+1)/\omega(k)$ and voltages $V(k+1)/V(k)$ in Eq.(3.2) we
obtained for the relative percentages $2.8{\%}$, $ 2.6{\%}$ and
$3{\%}$ for an average $2.5{\%}$ which demonstrates impressive
agreement with the resonance condition $2\pi X/a=\omega(k)$. Our
results confirm that when $V(k)/V_{0}$ is plotted versus the five $k$
values you obtain the function $[\omega
_{0}^{2}+4\sin^{2}(ka/2)]^{1/2}$ evaluated at the corresponding $k$
value as you should. The discreteness parameter in Fig.10, $\Lambda
=1.48$, is not sufficiently small enough to cause the discreteness
required to see the $j=2$ harmonics in $2\pi j\dot{X}/a=\omega(k)$.
However when Watanabe et al simulated the $I-V$ curve with $M=1$, $N=8$,
$\Gamma=.02$ and a smaller and thus more discrete $\Lambda ^{2}=1$
there were indications of two $j=2$ lines in their Fig. 11. The
voltage of the line they labeled $m=15$ is one half of the voltage of
the line $m=7$ and the voltage of their $m=9$ line is one half of the
$m=6$ line. The agreement of the ratios is less than three percent. If
the two lines $m=15$ and $m=9$ are $j=2$ lines than there are only
three $j=1$ lines instead of the 5 $j=1$ lines remaining. When we
assign the three remaining lines as follows $\omega(0)\leftrightarrow$
line 8, $\omega(\pi/4) \leftrightarrow $ line 7 and $\omega(3\pi/4)
\leftrightarrow $ line 6 their ratios compare with the simulation
ratios for $V_{8}$, $V_{7}$, and $V_{6}$ and voltages in Eq. (3.2) to
better than three percent. Furthermore the frequencies $\omega(3\pi
/4)$ and $\omega (\pi )$ are within six percent of each other and Fig.
11 is such that the $I$ and $V$ values are not carried out far enough
to be able to see the full $\omega(3\pi/4)$ and $\omega (\pi )$ lines.
Assuming that this is correct then we have 4 $j=1$ lines and 2 $j=2$
lines that agree well with the radiation resonance condition $2\pi
j\dot{X}=\omega(k)$.  There is one problem and that is the missing
$j=1$ line, the $\omega(\pi/2)$ line does not appear on the
numerically calculated I-V curve in Fig. 11.  Its absence is
unexplained and consequently all I can conclude is that we are
probably seeing 4 $j=1$ and 2 $j=2$ lines.

In addition to performing experiments on $N=8$ junction systems
Watanabe et al did a thorough study on an $N=4$ system with one kink,
$\Lambda ^{2}=1$ and $\Gamma =1$. The allowed $k$ values are $k=0,\pm
\pi/2$ and $\pi$ with frequencies $\omega(0)=1$, $\omega
(\pi/2)=1.73$, and $\omega (\pi )=2.24$. For $N=4$ there should be and
are three frequencies in their Fig. 9.  When we take the measured
voltages in Fig. 9 and form the ratios in Eq.(3.2) we obtain the first
ratio agrees to within two percent and the second ratio is essentially
exact so the average agreement with the phonon resonance condition is
one percent. The authors in [1] provide further insight into the above
$N=4$ system with the same parameters in their Fig.(8) by plotting
$\phi _{j}(t)$ vs. $t$ for each of the four junctions. From the figure
you can see that the $\phi _{j}(t)$ is a superposition of a kink and a
phonon with wavelength $\lambda=4a$, where $4a$ is the size of the
system, which means the $k$ of the phonon is $\pi/2$ because $ka =
\pi/2 \Rightarrow\lambda = 4a$. Consequently the velocity of the kink
is $\dot{X}/a = (2\pi )^{-1}\omega (\pi/2)= (2\pi )^{- 1} (3)^{1/2} $.
We can represent the kink and phonon configuration as $\leftarrow $x $
\to\to \leftarrow $ as it appears in Fig.8 where x is the center of
the kink and the arrows indicate the phase of the $k=\pi/2$ phonon
mode at the four junctions, and there is a change of $\phi $ of $2\pi
$ over the four junctions. Thus Fig. 8 is a beautiful conformation of
the fact that the kink and the phonon have exactly the same frequency
$2\pi\dot{X}/a=\omega(\pi/2)$ as they travel around the ring. Next we
consider the multiple kink case and find a completely new effect in
which the spacing of kinks in the ring allow only certain $k$ value
for the resonant phonons which we can explain by the CV radiation
theory.  The case of multiple kinks treated in Fig. 4 for the
parameters $N=8$, $\Lambda ^{2}=4$, $\Gamma $=.17, and for $M$=1,2,3
and 4 kinks. The value $\Lambda ^{2}$=4 (which is only weakly
discrete) corresponds to a kink size, 4$\pi a$ which is larger than
$8a$, the size of the system, which means all the kinks overlap each
other. In the continuum the SG is integrable and thus the kinks do not
interact with each other. In the present weakly discrete case the
discreteness is effectively a perturbation which destroys the
integrability and each kink interacts with all other kinks weakly
because they all overlap each other. It would be a difficult problem
to work out the solution of the equations of motion for $N$
interacting radiating kinks traveling around the ring.  However as we
see below we can infer the kinks travel around the ring keeping equal
spaces between each other. What is important is that the phonon modes
are independent of the number of kinks and are the same as the modes
in the absence of the kinks which is a consequence of the fact that
kink-phonon interaction in all cases in this paper can be treated
perturbatively. For the present case, $\Lambda ^{2}$=4, the kink
phonon interaction is sufficiently weak that only the lowest harmonic,
$j=1$, is observed. For $N=8$ as before there after five frequencies
that correspond to the $k$ values $k=0,\pm \pi/4a, \pm \pi/2a, \pm
3\pi/4a$, and $\pm \pi/a$.

We have as before $(V/V_{0})=(\omega/\Lambda)=[\Lambda ^{ -
  2}+4\sin^{2}(ka/2)]^{1/2} = [\omega_{0}^{2}+4\sin^{2}(ka/2)]^{1/2}=
(2\pi M\dot{X}/a\omega _{0})$. The fourth equality is the radiation
resonance condition for $M$ kinks. The only changes for $M \ge 2$ from
the $M=1$ case is that the phase change for $M$ kinks is equal to
$2\pi M, (2\pi )$ for each kink and the corresponding dimensionless
frequency is $2\pi MX/a\omega _{0}$, where all kinks have the same
velocity. The total rate of change of phase for a given $X$ is $M$
times the rate of change of one kink, i.e., $2\pi \dot{X}/a\omega
_{0}$.

In the experiment reported in Fig. 4 of [1] there are a total of four
lines one for each of the four values of $M$ i.e., $M=1,2,3,4$. When
we make the assignment $V(M)/V_{0} = [\omega _{0}^{2} +
4\sin^{2}(k_{M}a/2)]^{1/2}= 2\pi M\dot{X}_{M}/\omega _{0}a$ where
$k_{1}=\pi/4a, k_{2}=\pi/2a, k_{3}=3\pi/4a$ and $k_{4}=\pi /2a$ we
find the experimental values of $V(M+1)/V(M)$ agree with the
calculated frequency ratios $\omega(k_{M}+1)/\omega(k_{M})$ to within
two percent. Consequently the frequency assignments $\omega (M)$ are
clearly correct. The question arises why is there only one frequency
for each value of $M$ and why is that the frequency corresponding to
the assigned values $k_{2}=(\pi/2a), k_{3}=(3\pi/4a)$, and
$k_{4}=(\pi/a)$. We find the answer by finding the wavelength
$\lambda_{M}$ where $k_{M}=2\pi/\lambda _{M}$. For $M=2$
$\lambda_{2}=4a$, for $M=3$ $\lambda _{3}=8a/3$, and for $M=4$
$\lambda_{M}=2a$. Thus we find the wavelength of the resonant phonon
for each $M$ resonance is exactly the spacing between each other of
the $M$ kinks assuming the kinks are evenly spaced i.e., for $M=2$
kinks the spacing between the kinks is $4a$, $M=3$ the spacing is
$8\pi /3$ and $M=4$ the spacing is $2a$ in agreement with the $k_{M}$
values.  (Recall the total length of the ring is $8a$) The explanation
of the results are that the kinks are the source of radiation in the
linear inhomogenous wave equation for the phonons. We have already
seen that the frequencies in the source are the phonon frequencies
$2\pi MX/a$ for each $M$ but when we have $M\ge 2$ we also have a
spatial dependence in the source namely the equal spacing between the
kinks which determine the $k$ values of the phonons. That is the
spatial properties of the phonons radiated by the kinks are determined
by the spatial dependence of the source which is given by the
arrangement of the $M$ kinks rotating around the ring. The only
assumption is that for $M \ge 2$ the kinks moving around the ring are
evenly spaced. One would have to solve the full $M$ interacting kinks
and radiation problem to prove the assumption. Alternatively one could
reasonably conclude that in the solution the kinks were evenly spaced
by the agreement of the experimental results with the assumption of
equal spacing.

Up to point we have been using the radiation resonance condition to
determine the voltage at the line center of the line on the $I-V$
diagram. In the infinite discrete SG equation the phonon band is a
continuum and the radiation resonance condition $2\pi
j\dot{X}/a=\omega _{0}$ picks out a definite value of $X$ because the
phonon line is sharp. In the small system case we have a discrete,
damped and driven SG, Eq.(3.1), so that the valid perturbation theory
leads to a Lorentzian line shape for the phonon line which in turn
generates the step in the $I-V$ curves. There are two causes for the
linewidth in small discrete systems. First the damping term $\Gamma
\varphi _{j}$ gives a contribution to the linewidth proportional to
$\Gamma $ and contributes to each mode in the same way. The second
mechanism is the radiation linewidth due to spontaneous emission,
$\tau ^{ - 1}(k)$, which is different for each $k$ thus causing the
linewidth of $\omega (k)$ to depend on $k$. In addition, $\tau ^{ -
  1}(k)$ is a nonlinear function of $X$.  As a result, the delta
function that appears in the infinite system case $\delta[2\pi
j\dot{X}/a-\omega(k)]$ becomes a Lorentzian in the damped driven small
system, case

$$
[(2\pi j\dot{X}/a-\omega(k))^{2} + (\Gamma +\tau ^{ - 
1}(k))^{2}]^{-1} (\Gamma +\tau ^{ - 1}(k)) \eqno(3.3)
$$

The voltage $V$ is proportional to $X$ which is a nonlinear function
of $I$ through $\tau(k)$. Consequently in the small system $V$ is a
nonlinear function of $I$ which leads to the nonlinear $I-V$ curves in
the damped driven small system. The curves in the $I-V$ diagrams for
different $\omega(k$) thus have different widths and the lines from
adjacent $\omega(k)$ overlap each other leading to a step structure
and hysteresis. We explain the step structure and hysteresis in more
detail in the next section.

\section*{\large{IV. DISCUSSION AND CONCLUSIONS}}

In summary we have shown when the kink-phonon interaction can be
treated perturbatively that we can treat the discrete SG or (any
discrete non linear Klein-Gordon equation) as a ``point particle,''
$X(t)$ interacting with a gas of non interacting phonons where the
effective frequency of the kink is $\Omega=2\pi\dot{X}/a=K\dot{X}$ and
its harmonics $\Omega _{j}=j\Omega$ where $j$ is an integer. The
resonance conditions are $\Omega _{j}=\omega(k)$. In the large system
the phonon band is effectively a continuum and the resonance condition
appears as a delta function $\delta[\Omega _{j}-\omega(k)]$. In the
small system we have a Lorentzian with the same resonance condition
and a linewidth $(\Gamma +\tau ^{-1}(k))$.  For the perturbation
theory to be valid the system must be discrete enough for a moving
kink to produce radiation but not so discrete that the kink gets
trapped in a PN well or that nonlinear phonon-phonon interactions
become important.

In particular we have explained the experiments and simulations of a
small discrete periodic dynamically arrays of Josephson junctions in
[1]. Watanabe et. al [1] ended their paper by listing three open
problems requiring further research in the low voltage region of small
discrete Josephson junction arrays. In the present paper we have
solved all three problems.  Their first problem requiring solution was
can one find the correct resonance condition and derive it. The
correct answer which we obtained in this paper is, $2\pi
j\dot{X}/a=\Omega _{j}=\omega(k)$, i.e., the kink frequency is equal
to the phonon frequency which is the usual radiation resonance
condition that comes from the first Born approximation.

Their second problem to be answered was to find a precise
characterization of the steps in the low voltage region of the $I-V$
diagram. The steps in the $I-V$ diagrams are a direct consequence of
the kink phonon interactions. The separation of the phonon states in
the small Josephson junction rings is comparable or greater than their
width $\Gamma +\tau ^{ - 1}(k)$. Since $\tau $ depends on $k$ the
width of the steps are different for different $k$.  Consider the
driver $I$ driving the kink at a velocity $\dot{X}$ such that the kink
frequency $2\pi\dot{X}/a$ is in the region roughly between the phonon
frequencies $\omega(k-1)$ and $\omega(k)$. Then practically all the
energy from the driver goes into increasing the velocity from the
kink. As we increase $I$ then $\dot{X}$ (which is proportional to the
voltage $V$) increases until eventually $2\pi\dot{X}/a$ comes into the
linewidth of $\omega(k)$ and as the kink frequency approaches $\omega
(k)$ the kink begins to radiate. Thus $I$ has to increase rather
sharply to provide energy to the kink to compensate for the energy
loss of the kink due to its radiation of phonons and to prevent
$\dot{X}$ from decreasing.  Consequently there is a sharp increase in
$I$ for approximately the same $V$ i.e., $\dot{X}$.  As we increase
$I$ further $\dot{X}$ starts to increase thus it leaves the center of
the $\omega(k)$ line where the radiation from the $\omega(k)$ line
decreases to zero. Now as $I$ increases nearly all the energy from $I$
goes into again increasing $\dot{X}$ until the kink frequency enters
the wing of the $\omega(k+1)$ line and the whole process repeats
itself. Thus we have a precise qualitative and quantitative
characterization of the steps.

In the previous paragraph for convience I discussed the steps as if
the linewidths of $\omega(k)$ and $\omega(k+1)$ had a small overlap in
the wings of the lines. However the wings of the lines can overlap a
nontrivial amount and the explanation of the formation of the steps
remains valid but a new phenomena appears namely hysteresis. We can
understand the phenomena of hysteresis by the following argument. In
each step $\omega(k)$ the driver is a different nonlinear function of
voltage because $\tau(k)$ depends on both $k$ and $\dot{X}$. Or stated
differently the non linear function $I(V)$ is a different function for
each allowed $k$. When the linewidths of $\omega(k)$ and $\omega(k+1)$
overlap in their respective wings there are two stable states with the
same $\dot{X}$, i.e., voltage for a given $I$ corresponding to the
$\tau(k)$ of $\omega(k)$ and $\tau(k+1)$ of $\omega(k+1)$.  Starting
at the line center of $\omega(k)$ as $I$ is increased $\dot{X}$
increases and the $\omega $ linewidth becomes less stable and the
$\omega(k+1)$ linewidth becomes more stable until the stability of the
$\omega(k)$ vanishes and the system jumps to the $\omega(k+1)$ line.
Instead if you start at the line center of $\omega(k+1)$ with
linewidth $\tau(k+1)$ and decrease $I$ then $\dot{X}$ decreases and
eventually the system jumps discontinuously from the $\omega(k+1$)
line to the $\omega(k)$ line at a different value of $I$ (than the
increasing $I$ case) because $\tau(k+1)$ depends differently on
$\dot{X}$ then $\tau(k)$ does.

Their third problem requiring solution is ``can the prediction of the
high wave number resonances be experimentally observed?'' As we
discussed thoroughly in Sec. II, the resonance $2\pi j\dot{X}/a$ for
$j$=2,3..  are observable if the system is discrete enough i.e., if
$(\pi /l_{0})$ is sufficiently large or equivalently when $\Lambda $
is sufficiently small. The more discrete the system the more $j$
values that are observed. There is a maximum $j$ determined by the
discreteness that occurs when the kink becomes trapped in a PN well.

We conclude with the observation that the explicit solution for the
steady state radiation intensity and the explicit solution for
$\tau(k,X)$ can be obtained by solving for the steady state solution
of the linear Eq. (2.4) and calculating the radiation relaxation time,
$\tau $, in first order perturbation theory by the methods in [4] and
[7].

\newpage 
\begin{center}
{\bf REFERENCES}
\end{center}

\begin{enumerate}
 \item S. Watanabe, H.S.J. Vander Zant, S.H. Strogatz, and T.P.
  Orlando, Physica D {\bf 97}, 429 (1996) and references therein.
  
 \item R. Boesch, C.R. Willis and M. El-Batanouny, Phys. Rev.B {\bf 40},
  2284 (1989).

 \item M. Peyrard and M.D. Kruskal, Physica D {\bf 14}, 88 (1984).

 \item C.R. Willis, Phys. Rev. E {\bf 55}, 6097 (1997).
  
 \item C. Kittel, {\em Introduction to Solid State Physics}, Sixth Edition,
  John Wiley, N.Y., (1986).
  
 \item E.N. Economou, {\em Green's Functions in Quantum Physics},
  Second Edition, Springer-Verlag, Berlin, (1983).

 \item R. Boesch and C.R. Willis, Phys. Rev. B {\bf 42}, 2290 (1990).
\end{enumerate}

\end{document}